\documentstyle[12pt]{article}

\begin{document}
\begin{center}
{\bf Second Backbend in the Mass $A \sim 180$ Region}\\
\vspace*{0.5 true in}
A. Ansari$^{\rm a,b}$\\
$^{\rm a}$ Institute of Physics, University of Tokyo, Komaba\\
3-8-1 Komaba, Meguro-ku, Tokyo 153-08902, Japan\\
$^{\rm b}$ Institute of Physics, Bhubaneswar - 751 005, India\\
\end{center}
\begin{abstract}
\noindent Within the framework of selfconsistent cranked Hartree-Fock-
Bogoliubov theory(one-dimensional) we predict second backbend in the yrast
line of $^{182}Os$ at $I \approx 40 \hbar$, which is even sharper than the
first one observed experimentally at $I \approx 14 \hbar$.
 Around such a high spin the structure becomes multi-quasiparticle type,
but the main source of this strong discontinuity is a sudden large 
alignment of $i_{13/2}$  proton orbitals along the rotation axis
followed soon by the alignment of $j_{15/2}$ neutron orbitals. This leads
to drastic structural changes at such high spins. When experimentally
confirmed, this will be observed for the first time in this mass region,
and will be at the highest spin so far.
\end{abstract}

\vspace*{0.3 true in}


\newpage
    During the latter half of the seventies and early eighties when the
research activities in the high spin spectroscopy were rather at its peak,
experimentally as well as theoretically, it was the one-dimensional
cranking (rotation about a principal axis(x) perpendicular to the
symmetry axis(z)) Hartree-Fock-Bogoliubov (CHFB) theory \cite{Ring npa215} - 
\cite{Dieb npa345}
which was most successful (besides its well-known shortcomings) to explain
the underlying mechanism of the backbending features observed in the
moment of inertia (or spin) versus rotational frequency plot of the
yrast levels (level with the lowest energy for a given angular momentum). This
is because of the fact that the mechanism of the alignment of single particle
 angular momenta along the rotation axis(the effect of collective rotation
on single particle motion) is naturally present in this approach.

 After exhaustive activities in the rare-earth region, now the high spin
 structure work has spread to all the mass regions of the periodic table
 with the maximum spin value reached being of the order of $I = 50 \hbar $(it
is $60 \hbar$ for the first superdeformed band in $^{152}Dy$).
Besides the yrast sequence, several side bands are observed in most
of the nuclei. Recently very interesting features have been observed in
some of the nuclei near the upper end of the rare-earth region, namely, 
in $ W - Os $ isotopes with mass number $A \sim 180$ \cite {Kuts npa587}
 - \cite{Fire tab96}. In some even - even
isotopes side bands are found with high-K band heads (K being projection
of I along the z-axis) very close to the yrast line with signature
 ( $r = (-1)^I$ ) symmetry broken, that is, even and odd spin states are
connected by B(M1) transitions and the ratio $B(M1)/B(E2)$ is found to be
large. Such states with mixed signature symmetry are interpreted as
t bands \cite{Frau npa557} arising due to the rotation of the nucleus about an
 axis tilted 
with respect to the principal axes of the quadrupole shaped deformed nucleus.
Some of the states of these high-K bands decay to the yrast states,
particularly in the band crossing region. So, it becomes natural to expect that
the mutual interaction between these unperturbed bands should influence
the backbending behaviour in the yrast sequence. Also in this mass region
the Fermi surface lies in the high-m states of high-j orbitals(e.g. $0h_{11/2}$
for protons and $0i_{13/2}$ for neutrons) which also gives some credence
to such expectation that the yrast states, at least in the band crossing
region, could be generated by tilted axis rotation. Thus, it has become
an interesting and challenging problem to test if the usual explanation
of backbending caused by the crossing of the s band ( low-m $i_{13/2}$
neutron aligned band) still holds. This question is not yet resolved
through a microscopic quantal many body calculation. It may be added that
 currently Onishi
and his collaborators \cite{{Hori npa596},{Moi prep98}} are attempting
to perform a generator coordinate method (GCM) calculation for $^{182}Os$ 
 treating the tilting angles as generator coordinates. The good angular
 momentum projected GCM wave functions would be able to elucidate on the
distribution of K components for a given value of the angular momentum.

In some odd-A nuclei in this mass region it has recently been found that
the high-K bands really cross the ground band producing backbend in
the yrast sequence,e.g. $^{179}W$ \cite{Walk npa568} and
 $^{181}Re$ \cite{Pear prl79}. 
However, for even - even nuclei the general conclusion so far is that
it is still the normal s band crossing that plays essential role to
produce the backbend in these nuclei \cite{Shiz npa593}.
Furthermore, in a recent microscopic theoreical analysis of the tilted axis
rotation following the band mixing spin projected shell model approach  
\cite{Sheik prc57}
it is found that in the case of $^{178,180,182}W$ and $^{184}Os$ the backbend
originates due to normal s band crossing; the tilted bands appear slightly
above the yrast line. In view of these findings  we thought of checking
as to how the traditional CHFB approach works for this mass region.  
Particularly for very high spins, $I > 20 \hbar $, the usual CHFB theory is
expected to work well, and angular momentum projection is almost impracticable.
 Hence, we have carried out here a selfconsistent CHFB calculation 
for $^{182}Os$ in an appropriate single paricle model space.
For this nucleus presently the yrast line extends up to $I=34 \hbar$ 
 \cite{{Kuts npa587},{Fire tab96}}
without a second backbend and the calculation has been performed for spins up
 to $I = 50 \hbar$. We find a strong backbend
at around $I = 40 \hbar$. In the followings we present our results and
discussions on these. Also now on the spin values will be understood to be in
the units of $\hbar$.

  Since the CHFB theory \cite{{Ring mb80},{Good adv11}} is well-known we
will not give any details here. For the Hamiltonian of the system we have used 
a pairing-plus-quadrupole model interaction. However, in addition, 
a hexadecapole term is also considered as these nuclei are expected to have
large (negative) hexadecaploe deformation, $\beta_4$ \cite{Naz npa512}.
The total Hamilonian can be written as
\begin{equation}
\hat H = \hat H_0 - \frac{1}{2}\sum_{\lambda=2,4}\chi_\lambda \sum_\mu
\hat Q_{ \lambda \mu}(-1)^\mu \hat Q_{\lambda -\mu} -
\frac{1}{4} \sum_{\tau = p,n} G_\tau \hat P^\dagger_\tau \hat P_\tau
\label{H}
\end{equation} 
 where $\chi_\lambda$ and $G_\tau$ are the corresponding interaction strengths
(in MeV), and the multipole moments ($\hat Q_{\lambda \mu}$) and the pairing
 operator ($\hat P$) have standard forms \cite{Ring mb80}, the radial part of
the former being ($r^2/b^2$) with $b$ as the oscillator length parameter.
$\hat H_0$ represents the spherical part of the hamiltonian with the single
particle (s.p.) orbitals(assuming $Z=40, N=70$ core) $2s_{1/2}$, $1d_{3/2}$,
$1d_{5/2}$, $0g_{7/2}$, $0g_{9/2}$, $0i_{13/2}$, $1f_{7/2}$, $0h_{9/2}$, and
$0h_{11/2}$ for protons, and $2p_{1/2}$, $2p_{3/2}$, $1f_{5/2}$, $1f_{7/2}$,
$0h_{9/2}$, $0h_{11/2}$, $0j_{15/2}$, $1g_{9/2}$, $0i_{11/2}$, and $0i_{13/2}$
 for neutrons. The s.p. energies
are as given in table.1 of Ref. \cite{Hori npa596} except that of $0j_{15/2}$
 which is 7.179 MeV. These energies are essentially the Nilsson spherical
s.p. energies for this mass region. 

As indicated above, Eq. (\ref{H}) actually represents two Hamiltonians as
far numerical calculations are concerned here. One is as it is, and in the
other the hexadecapole term is dropped. Correspondingly we have two sets of
ineraction strengths which can reproduce more or less the ground state
intrinsic shape parameters of $^{182}Os$.
Finally taking principal x-axis as the cranking axis the CHFB eigenvalue  
equations are solved by diagonalization with the usual particle number and
angular momentum constraints \cite{{Ring mb80},{Good adv11}}.

 We have also computed rotational g Factors using the 
standard cranking expression \cite{Ans npa415}
\begin{equation}
g_I = \langle \hat \mu_x \rangle / \sqrt{I(I+1)}
\end{equation}
where $\hat \mu_x$ is the x-component of the magnetic moment operator
\begin{equation}
\hat \mu = g_l \sum_i \hat j_x(i) + \left ( g_s - g_l \right ) \sum_i
 \hat s_x(i)
\end{equation}
with $g_l = 1$ and $g_s = 5.586$ for protons and $g_l=0$ and $g_s = - 3.826$
for neutrons. In numerical computation the values of $g_s$ are attenuated
by a factor of 0.6 \cite{{Ans npa415},{Walk npa568}}.

The CHFB equations are solved selfconsistently in terms of seven collective
variables when the hexadecapole(Q4) term is also present in the Hamiltonian.
These are the pairing gaps, $\Delta_p$ and $\Delta_n$, for protons and 
neutrons, respectively and deformation parameters $q_{\lambda \mu}=
 q_{\lambda -\mu} =
\langle \hat Q_{\lambda \mu} \rangle $ with $\mu = 0, 2$ for $\lambda = 2$ and
$\mu = 0, 2, 4$ for $\lambda = 4$.  The usual deformation parameters ( $\beta$,
 $\gamma$ ) and $\beta_4$ are defined through the relations,
$\hbar \omega_0 \beta \cos \gamma = \chi_2 q_{20}$,
$\hbar \omega_0 \sin \gamma / \sqrt{2} = \chi_2 q_{22}$ and
$\hbar \omega_0 \beta_4 = \chi_4 q_{40}$,
 where $\hbar \omega_0 = 41/A^{1/3}$ MeV.
 
When only quadrupole interaction is considered then the degree of freedom is 
reduced to four. As already mentioned, the interaction strengths are chosen 
such that the values of the ground state shape parameters (see Table I)
are approximately reproduced  \cite{{Kuts npa587},{Naz npa512}}.
The values of the pairing gap parameters are decided by looking at the 
experimental odd-even mass differences. It may be emphasized that after fixing
the interaction strength parameters at this stage there are no
free parameters in the theory. In the followings, for the sake of brevity,
when the quqdrupole + hexadecapole both the interaction terms are considered
, then the corresponding results will be indicated by the symbol "Q4".
But if only quadrupole term is considered, then it will accordingly be 
indicated by "Q2".

Now we can discuss some of our main results. In Table I we have listed
the values of the shape parameters at a few angular momentum values, the
dependence on $I$ being most striking around $I = 40$. In both the cases,
with and without the Q4 term, $\Delta_n$ goes to zero at $I = 14 $.
On the other hand $\Delta_p$ vanishes at $I = 30$ in presence of the Q4 term,
and at $I = 26$ without it which implies a somewhat  stronger proton pairing
correlation in the former case for the spin range of about $I = 16 - 28$.
We find that when only quadrupole force is considered
the proton pairing recovers for $I = 34$ ($\Delta_p = 0.214$ MeV) to
$I = 40$ ($\Delta_p = 0.239$ MeV). In the other case it recovers only
at one spin, $I = 34$ with a small value of $\Delta_p = 0.174$ MeV.
In both the cases there is a correlation between $\gamma $ acquiring a 
negative value and an increase in the value of $\beta $. Around spin $I = 40$ 
 $\gamma$ changes sign from positive to negative by a quite sizeable amount,
 and associated with it the value of $\beta$ increases by about $20 \%$
(stretching). At the same time the value of $\beta_4$ shows a sudden 
decrease. However, it may be pointed out that now all the three
components of $q_{4 \mu}$, $\mu = 0, 2, 4$ become of similar magnitude: for
instance, at $I = 38$ these are -8.05, -1.09 and 1.08 which become -4.10, -2.36
and 2.56 (all in units of $b^2$) at $I = 40$ for $\mu = $ 0, 2 and 4, 
respectively.

 In Fig. \ref{i-w} we display a backbending(BB) plot of spin versus rotational
frequency ($\omega$) where for the experimental case 
$\omega_I = \frac{1}{2} (E_I - E_{I-2})$. As indicated on the top right corner
of the figure, the three curves correspond to the experimental data, with
hexadecapole(Q4) and without hexadecapole(Q2) terms in the Hamiltonian of
 Eq. (\ref{H}).
We notice that the first backbend is actually not well reproduced, though an
upbend is produced at more or less the correct frequency, and the inclusion
of the hexadecapole degrees of freedom helps in the right direction.
However, the alignment of
$0i_{13/2}$ neutron orbitals is quite pronounced as can be seen in the next
 Fig. \ref{ali}, where contributions from a few important orbitals
to the total angular momentum are shown for the "Q4" case.
 At $I = 14$ the contribution of $ni_{13/2}$ orbitals (mainly $m = 7/2, 9/2$
components) is about $9 \hbar$, close to the experimental estimate of 
$10 \hbar$ \cite{Kuts npa587}. However, the alignment
is not sudden enough around this spin to cause a sharp BB. Hence, we may
conclude that some extra mechanism is, perhaps, needed to obtain a sharp
first BB. We are trying to perform angular 
 momentum projection, including K-mixing, on CHFB wave functions. If this
also fails, then, perhaps, tilting mechanism is the only explanation.          
   
However, the main interesting result here is the appearance of a second BB near
$I = 40$ as seen in Fig. \ref{i-w}. The small backward kink in the "Q4" curve
 at $I = 36$ is a genuine one, that is, it is not due to some numerical
inaccuracy etc. The sharp BB is, of course, due to a sudden large and
coherent alignment of the $pi_{13/2}$, $ph_{9/2}$ and $nj_{15/2}$
 orbitals (mainly $m$ = 1/2 and 3/2 components) as clearly seen in
Fig. \ref{ali}. At $I = 40$ the contribution of $ni_{13/2}$ suddenly
drops by about 6 units in one step. The $nh_{9/2}$ orbitals contribute
at all spins, in almost a gradual manner, whereas $ph_{11/2}$ orbitals
start contributing at very high spins through high-m components.
Thus, the structure near $I = 40$ is very interesting. 
There is a quite sizeable stretching of $\beta$, and $\gamma$ acquires
a negative value. That is in this BB region collectivity increases, rather
than showing a decrease, as is often observed in the first band crossing
region \cite{{Oshima prc33},{Mnr prl57}}.
 For the spin region $I = 40 - 50$ the intrinsic structure remains
essentially unchanged, which may be seen as a second minimum in a shell
correction calculation, and levels for $I \geq 40$ may be interpreted as
rotational levels in the second well. The increase in $\beta$ can be
understood as due to an enhanced magnitude of the quadrupole matrix elements
of the low-m, high-j (aligned) neutron and proton orbitals near the
Fermi surface.
 
In Fig. \ref{g-f} is displayed a variation of the g Factors with spin. The 
actual value of $g(I=2)$ is 0.245 for the "Q4" case and 0.230 for the "Q2"
case which appear to be reasonable in view of $g_R = 0.27$ used in
Ref. \cite{Kuts npa587}.
In both the cases the ratio $g(I)/g(2)$ drops sharply to a minimum at
$I = 14$, though in "Q4" case the real minimum is at $I = 20$ with a slightly
lower value. The sharp 
rise at $I = 40$ is a clear indication of the large alignment of the proton
orbitals. Beyond this the alignment of $nj_{15/2}$ orbitals stops the further 
rise. In the intermediate spin region ($I = 20 - 30$) relatively larger 
magnitude of the alignment of the large-m components of $ph_{11/2}$ make
 the g Factors higher for the "Q2" case.

 Finally we would like to make some additional remarks. Since at $I= 40$
the pairing has collapsed in our calculation, one may think that in a particle
number projected treatment the position of BB may get shifted or become
much less dramatic. In order to check for this a calculation was performed
for $I > 30$ with a fixed value of $\Delta_p$ and $\Delta_n$ at about
half of their values in the ground state. Then it is found that
$\gamma$ changes sign between $I = 32$ to 34 with all the
characteristics as noted above. Thus, the second BB seen here seems
very much genuine. We also notice a small favourable trend looking at the
difference of the experimental $\gamma$ - ray energies:  
$E_\gamma (32)$ - $E_\gamma (30)$ = 78 keV, and $E_\gamma (34)$ 
- $E_\gamma (32)$ = 65 keV.

In conclusion, through a selfconsistent CHFB calculation, which is very
reliable at high spins ($I \geq 20$), we have obtained  a clear case
of sharp second backbend in $^{182}Os$ near $I = 40$. This is caused by 
a large coherent alignment of low-m $pi_{13/2}$, $ph_{9/2}$ and $nj_{15/2}$
orbitals. In this spin region there is a substantial change of structure
within a couple of units of angular momentum. For instance, $\gamma$ goes
positive to negative, with a change of $12^o$ or more in one step, 
with an associated increase of $\beta$ by about $20 \% $(stretching).
However, the first backbend is not well reproduced, and, as discussed above,
we are working on it. We have also studied the variation of $g$ Factors as a
function of spin which essentially shows, in a gross manner, the alignment
pattern of neutron and proton single particle orbitals.

 Particularly for $^{182}Os$ the levels are already known up to $I = 34$,
 so we hope that our prediction will produce enough excitement in
experimentalists to put efforts to study the interesting features in the spin
$I = 40$ region.

The author is grateful to Naoki Onishi for his kind support and many useful
discussions. He would also like to acknowledge the financial support from
the Japan Society for the Promotion of Science.

\newpage   
\begin{table}
\caption {Intrinsic shape parameters as a function of spin, I (actually
$\sqrt{I(I+1)}$). For the ground state $\gamma = 0$ and $\beta, \beta_4,
\Delta_p$ and $\Delta_n$ are, respectively 0.228, -0.038, 0.871 MeV, and
0.879 MeV in case "Q4", and 0.229, 0.0, 0.872 MeV and 0.886 MeV in case
"Q2". At I = 10 $\Delta_n = 0.436$ MeV in the former case and 0.432 MeV
in the latter case. It goes to zero for $I \geq 14$.}
\begin{tabular}{|c c c c c |c c c |}
\hline
\multicolumn{5}{|c|}{with hexadecapole}&
\multicolumn{3}{c|}{without hexadecapole}\\
\multicolumn{1}{|c}{I}&
\multicolumn{1}{c}{$\beta$}&
\multicolumn{1}{c}{$\gamma$}&
\multicolumn{1}{c}{$\beta_4$}&
\multicolumn{1}{c|}{$\Delta_p$}&
\multicolumn{1}{c}{$\beta$}&
\multicolumn{1}{c}{$\gamma$}&
\multicolumn{1}{c|}{$\Delta_p$}\\
\multicolumn{1}{|c}{( $\hbar$ )}&
\multicolumn{1}{c}{ }&
\multicolumn{1}{c}{(deg)}&
\multicolumn{1}{c}{ }&
\multicolumn{1}{c|}{(MeV)}&
\multicolumn{1}{c}{ }&
\multicolumn{1}{c}{(deg)}&
\multicolumn{1}{c|}{(MeV)}\\
\hline
0 & 0.228 & 0.0 & -0.038 & 0.871 & 0.229 & 0.0 & 0.872 \\
10& 0.235 & 1.58& -0.039 & 0.799 & 0.237 & 1.35 &0.796 \\
20& 0.224 & 5.89& -0.047 & 0.699 & 0.227 & 4.11 &0.571 \\
30& 0.207 & 10.58&-0.050 & 0.0   & 0.208& 9.53 & 0.0 \\
38 &0.217 & 0.97& -0.050 & 0.0   & 0.203 &8.62 & 0.262 \\
40 &0.253 &-11.11&-0.025 & 0.0   &0.209 &4.88 & 0.230\\
42 &0.256 &-12.80&-0.020 & 0.0   &0.260 &-12.52& 0.0 \\
50& 0.245& -12.70&-0.021 & 0.0   &0.253& -12.63& 0.0\\
\end{tabular}
\end{table}
\begin{figure}
\noindent{\bf Figure Captions:}\\
\caption{\label{i-w} Backbend plot for $^{182}Os$ showing the variation of
angular momentum with rotational frequency, $\omega$. As indicated on the 
figure the curve with solid line corresponds to the experimental data with
$\omega_I = \frac{1}{2} (E_I - E_{I-2})$. The long dashed curve, labeled "Q4"
corresponds to the case when quadrupole and hexadecapole both the interaction
terms are present in the Hamiltonian (\ref{H}). The short dashed curve,
labeled "Q2" indicates that only quadrupole interaction is considered.}
              
\caption{\label{ali} Alignment plot. Contributions of a few important orbitals
to the total spin is displayed denoting it as a single particle(S.P.)
contribution. These are corresponding to the "Q4" case only. The type of 
orbitals is indicated for each curve, where p and n indicate proton and  
neutron, respectively.}
         
\caption{\label{g-f} Variation of the g Factor ratio $g(I)/g(2)$ as a 
function of spin for both the cases, that is, with the inclusion of
hexadecapole term(Q4), and without it(Q2).}
\end{figure}

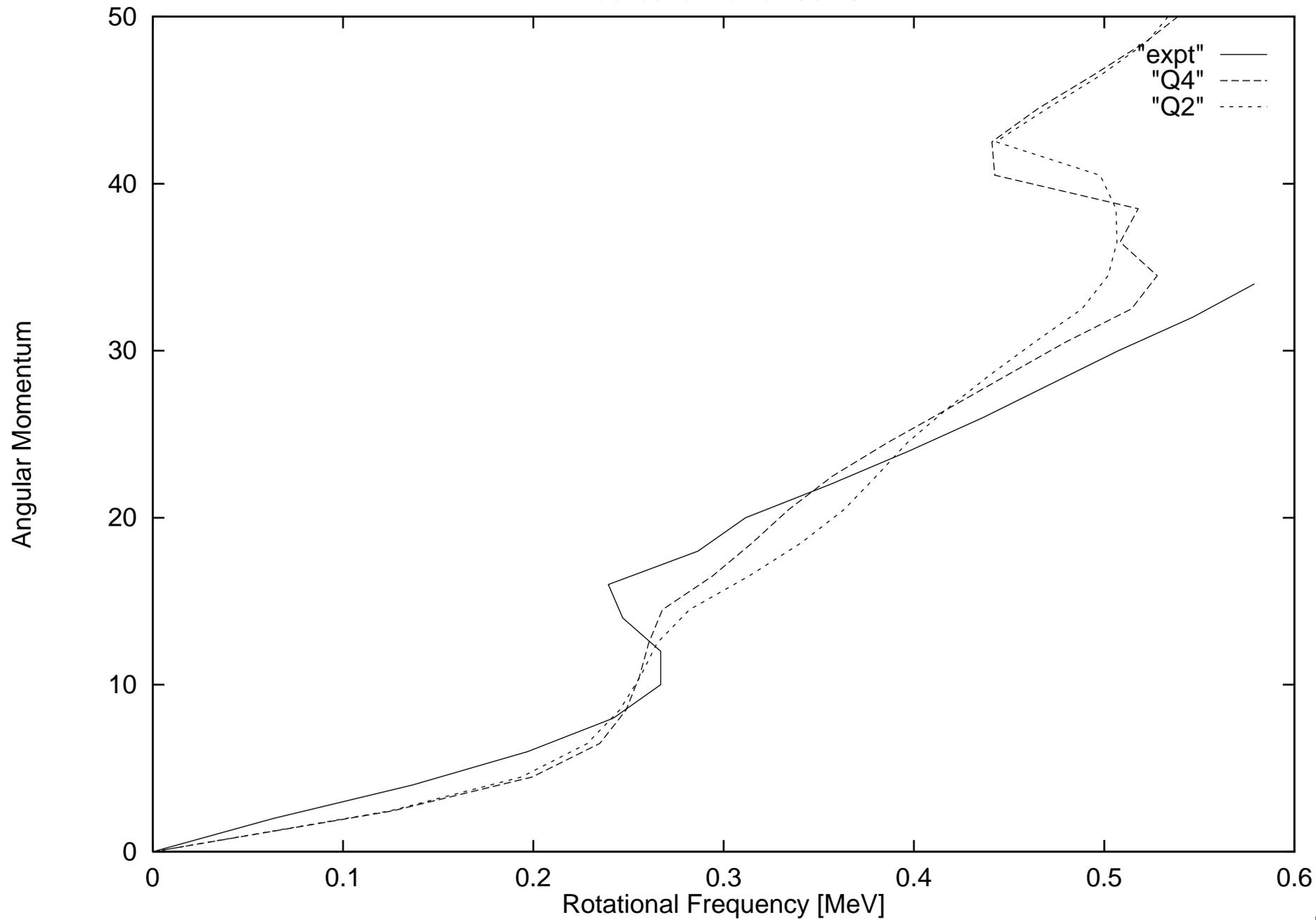
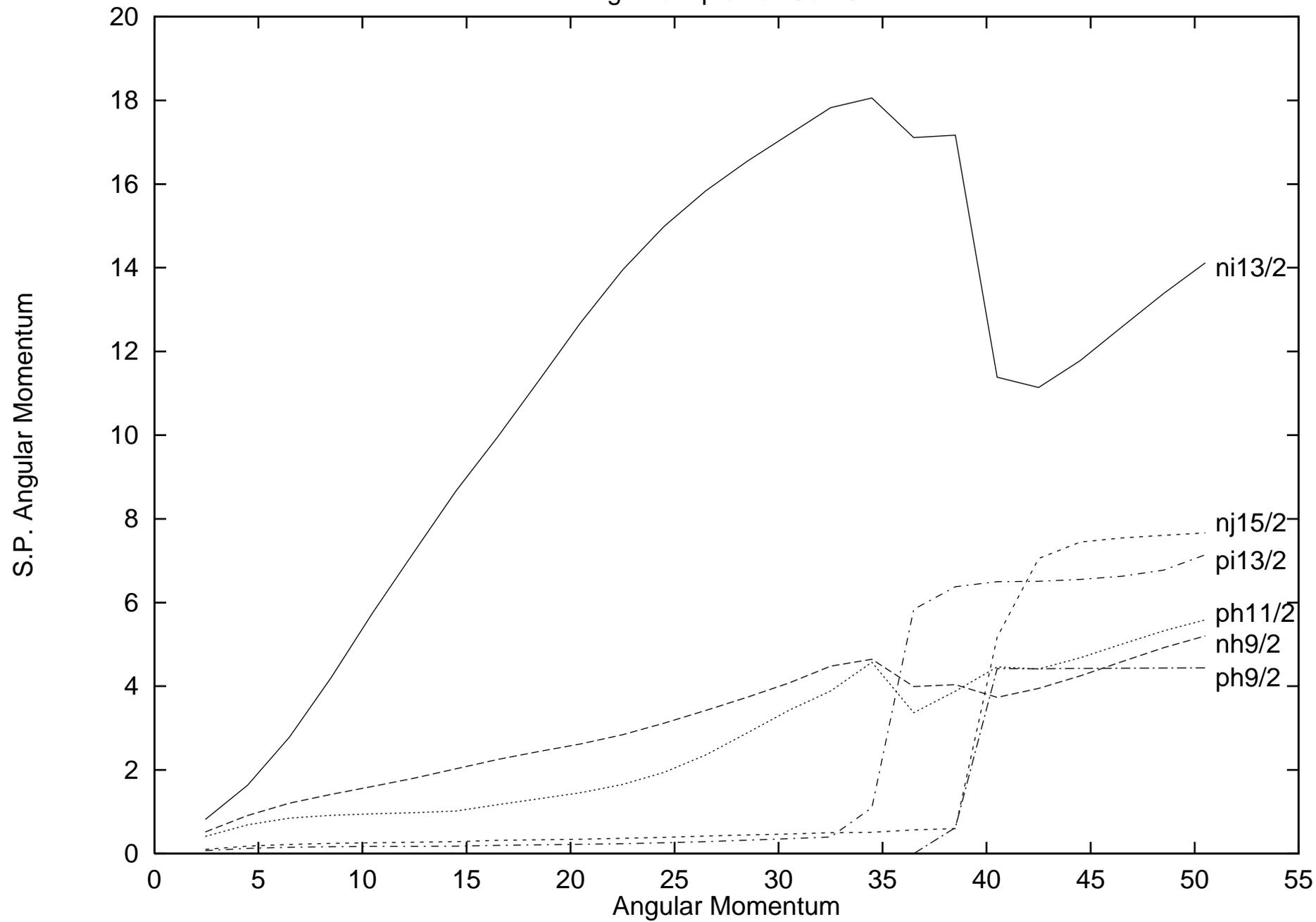
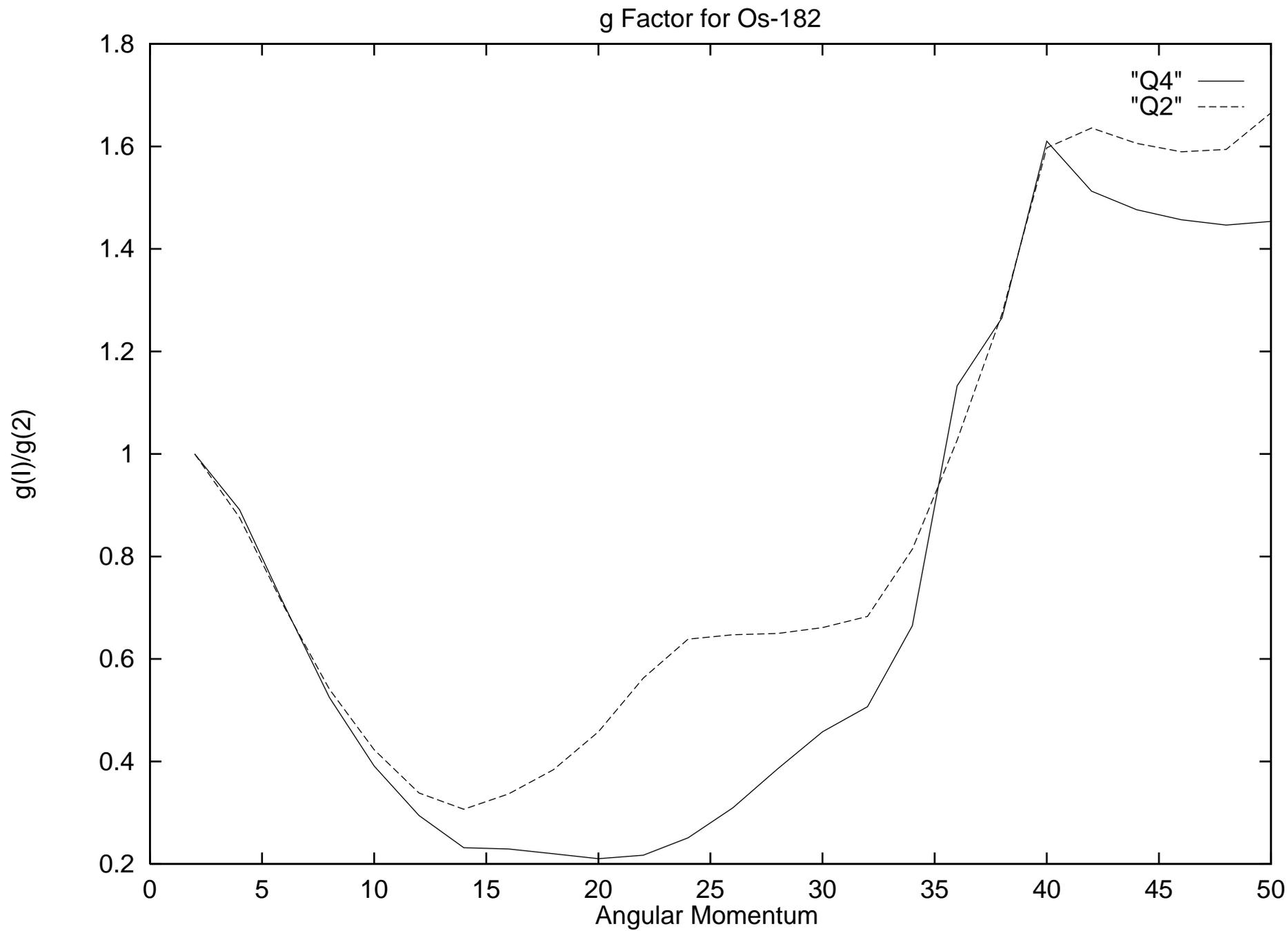
\end{document}